\documentclass{iau} 
\usepackage{graphicx}
\usepackage{xspace, amsmath}
\usepackage{natbib}
\usepackage{tikz}
\usepackage{hyperref}
\usetikzlibrary{shapes, arrows}

\newcommand{\D}{\mathrm{d}}

\newcommand{\mnras}{\textit{MNRAS}}
\newcommand{\aap}{\textit{Astron. \& Astrophy.}}
\newcommand{\apj}{\textit{ApJ}}
\newcommand{\raa}{\textit{Research in Astron. and Astrophy.}}

\graphicspath{{figures/}}

\newcommand{\mytab}{\begin{table}[htb]}
\newcommand{\myfig}{\begin{figure}[htbp]}

\newcommand{\msun}{\mathrm{M}_\odot}

\title[satellite tracer]{Satellite galaxies as better tracers of the Milky Way halo mass}
\author[J. Han et al.]   
{Jiaxin Han$^{1,2}$,
 Wenting Wang$^{2,1}$
 \and Zhaozhou Li$^{1}$}

\affiliation{$^1$Department of Astronomy, Shanghai Jiao Tong University, Shanghai 200240, China \\[\affilskip]
$^2$Kavli IPMU (WPI), UTIAS, The University of Tokyo, Kashiwa, Chiba 277-8583, Japan
\\ email: {\tt jiaxin.han@sjtu.edu.cn}
}

\pubyear{2019}
\volume{353}  
\jname{Galactic Dynamics in the Era of Large Surveys}
\editors{M. Valluri, \& J. A. Sellwood  }
\begin{document}

\maketitle

\begin{abstract}
The inference of the Milky Way halo mass requires modelling the phase space structure of dynamical tracers, with different tracers following different models and having different levels of sensitivity to the halo mass. For steady-state models, phase correlations among tracer particles lead to an irreducible stochastic bias. This bias is small for satellite galaxies and dark matter particles, but as large as a factor of 2 for halo stars. This is consistent with the picture that satellite galaxies closely trace the underlying phase space distribution of dark matter particles, while halo stars are less phase-mixed. As a result, the use of only $\sim 100$ satellite galaxies can achieve a significantly higher accuracy than that achievable with a much larger sample of halo stars.
\keywords{dark matter -- galaxies: haloes -- methods: data analysis}
\end{abstract}

\firstsection 
\section{Introduction}
The outer halo potential of the Milky Way have been modelled with a variety of methods, leading to a range of results with typical values varying from $\sim 0.5\times 10^{12}\msun$ to $\sim 2\times 10^{12}\msun$ (see \citealp{Wang15}; Wang et al. 2019 in prep, for compilations of recent measurements). While part of this divergence can be attributed to statistical uncertainties, systematic biases must also play a significant role given the small errorbars in some of the measurements. Understanding such systematics is becoming more and more important as ongoing and upcoming surveys of the Milky Way such as LAMOST~\citep{LAMOST}, GAIA~\citep{GAIA} and 4MOST~\citep{4MOST} are collecting more complete and accurate phase space data on halo stars, globular clusters and satellite galaxies. These data will further reduce the statistical uncertainty of dynamical models and as we show below, lead to systematics dominated results in some of the analysis.

This work aims at understanding and quantifying the systematics associated with a large family of models that assume the tracer distribution is in a steady-state. It has been shown that unjustified assumptions in such modelling can lead to significant biases~\citep{Wang15,oPDFII}. Therefore, we will use a minimal assumption method that avoids any unnecessary assumptions, in order to understand the most fundamental systematics associated with the steady-state assumption. This approach also makes our results to have wide implications for the entire family of steady-state methods~\citep{Wang18}. We test the method on three types of tracers, including dark matter particles, halo stars and satellite galaxies. The mock data used are extracted from simulations in the $\Lambda$CDM cosmology, including the APOSTLE hydrodynamical simulations of Local Group like systems~\citep{Apostle}, and the Millennium-II $N$-body simulation~\citep{MRII} from which we extract a large sample of Milky Way like haloes. 

\section{Minimal assumption steady-state modelling}
We use the orbital Probability Distribution Function~\citep[oPDF;][]{oPDF} method as a benchmark method for steady-state modelling. The starting point of the method is that in a steady-state system, along a given orbit,
the probability density of tracer particles is proportional to the travel time at each position. 
In a spherical system, specifying the orbits by the energy and angular momentum parameters, the above principle can be explicitly written as
\begin{equation}
\D P(r|E,L)=\frac{\D r}{v_r(E,L,r) T(E,L)},\label{eq:oPDF}
\end{equation} where $v_r$ is the radial velocity and $T$ is the radial period.

Equation~\ref{eq:oPDF} is sufficient for the inference of the halo potential. Starting from a trial potential $\psi(r)$, one can convert the observed phase space coordinates of tracers to their orbital parameters $(E,L)$. The distribution of orbital parameters coupled with Eq.~\ref{eq:oPDF} can be used to predict an overall spatial distribution $P_\psi(r)$. Finally, the potential can be constrained by comparing the predicted and observed spatial distributions in a likelihood framework. Note that the only essential assumption of the method (besides spherical symmetry) is that the tracers are in a steady-state. In the following, we will apply the method while parametrising the halo potential with a total halo mass, $M$, and a concentration, $c$, following an Navarro-Frenk-White~\citep{NFW97} profile. 

\begin{figure}
\includegraphics[width=0.8\textwidth]{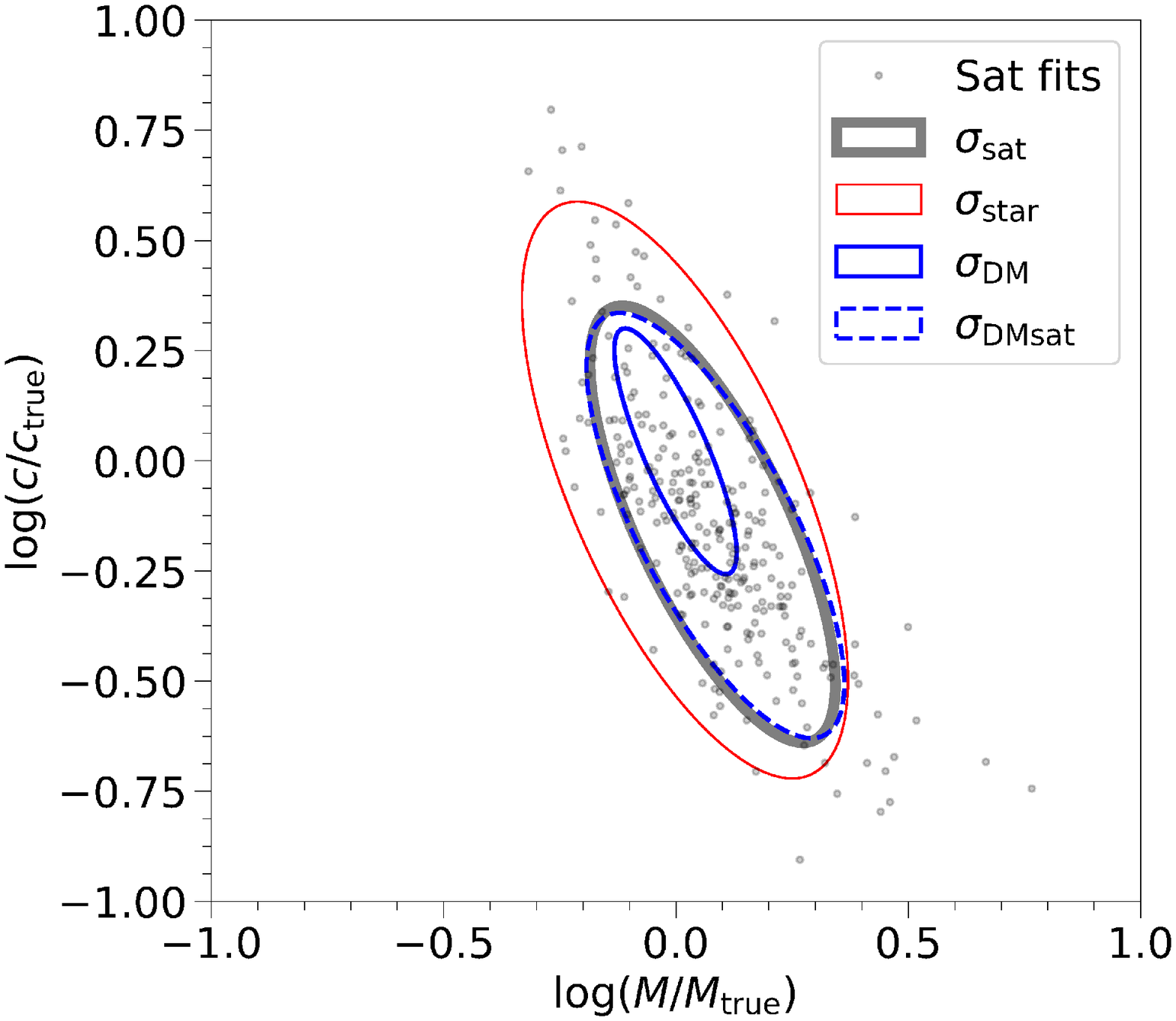}
\caption{Fits to different tracers in MW size haloes using the oPDF method. The grey points show the fits to individual haloes using satellites as tracers. 
The thick grey ellipse ($\sigma_{\rm DM}$) shows the total $1-\sigma$ scatter of the data points. The red ($\sigma_{\rm star}$) and blue ($\sigma_{\rm DM}$) solid ellipses show the total scatters in the fits using stars and dark matter particles as tracers respectively~\citep{Wang17}. The blue dashed ellipse ($\sigma_{\rm DMsat}$) shows the total scatter associated with dark matter particles down-sampled to the have the same number as that of satellites.
}\label{fig:tracers_comp}
\end{figure}

Applying this method to a large sample of Milky Way like haloes selected from the Millennium-II simulation, \citet[][hereafter Wang17]{Wang17} found significant stochastic biases in the fitted halo parameters. These biases are caused by correlated phase space structures in the tracers that violate the steady-state assumption. The effect of phase-correlation is to reduce the number of independent tracers. As a result, the deviations of the fits from the true values can be much larger than that allowed by the statistical uncertainty derived from the apparent tracer sample size. 

The amount of phase-correlation is determined by the merger history of the halo, and cannot be reduced by simply increasing the number of tracer particles. These irreducible stochastic biases represent the limit of steady-state information that can be extracted from the tracers. Using dark matter particles as tracers, Wang17 found this stochastic bias to be typically $\sim 20\%$ in mass. Similar stochastic biases are found using halo stars in the APOSTLE hydrodynamical simulation, but at a much higher level of $\sim 100\%$ (see Fig.~\ref{fig:tracers_comp}). These correspond to an effective number of phase-independent particles of $\sim 1000$ in dark matter particles and only around $40$ in halo stars. 



\section{Satellite galaxies as better steady-state tracers}
Given the small effective number of phase-independent particles in halo stars, better measurements of the halo mass can be expected if one can find a tracer with a larger number of phase-independent particles. Indeed satellite galaxies can serve as such a tracer. To show this, we apply the oPDF method to satellite galaxies in the Milky-Way like halo sample of Wang17 selected from the Millennium-II simulation. 
Wang17 have analysed both isolated and binary haloes. The differences between the two is subdominant compared to the differences between different tracers, so we only focus on the set of isolated haloes here. The satellites are extracted from the semi-analytic galaxy catalogue of \citet{Guo11} in the Millennium database.\footnote{http://gavo.mpa-garching.mpg.de/Millennium/} Our results are not sensitive to how the satellite galaxies are selected. We have tested that using a peak-mass limited satellite subhalo sample from a different subhalo finder~\citep[HBT+;][]{HBTplus} leads to very similar results.

The results are shown in Fig.~\ref{fig:tracers_comp}. To make a consistent comparison with Wang17, we apply the same radial cut of $20{\rm kpc}<r<R_{200}$ when applying oPDF to satellite galaxies. In addition, to decrease confusions from statistical noise, we only use haloes with more than 100 satellites in the given radial range. {Selecting haloes according to the number of satellites are not expected to bias our results~\citep{Li19}.} 

The ellipses in Fig.~\ref{fig:tracers_comp} show the total uncertainty (statistical plus systematic) associated with each type of tracer. Note that the number of tracer particles used in Wang17 are around $10^5$, leading to negligible statistical uncertainties in the star and dark matter tracers. On the other hand, the number of satellites in each halo available in our analysis is $\sim 100$, corresponding to a much larger statistical uncertainty. Despite this, the total uncertainty in stars is still much larger than that in satellites. This means the systematic uncertainty in stars has to be the highest among the three tracers. 

To further compare the systematic uncertainties between satellite and dark matter particles, we down-sample the dark matter tracers to have the same number of particles as the number of satellites. This ensures the two tracers to have comparable statistical uncertainties, so that the differences between their total uncertainties can be attributed to their different systematics. Interestingly, as shown in Fig.~\ref{fig:tracers_comp}, the total uncertainties between the two are almost identical. This means the dynamical state of satellite galaxies is close to that of dark matter particles. 

This may not be surprising as satellites, especially small ones, are not much different from dark matter particles with a large mass. This is also consistent with previous models of the spatial distribution of satellites in which satellites are assumed to sample the phase space distribution and accretion of dark matter particles in a \emph{unbiased} way~\citep{Han16}. By contrast, halo stars are stripped along correlated orbits off satellite galaxies, leaving a high degree of phase-correlation among them and thus stronger deviations from steady state. Although satellite galaxies can also be accreted in groups, the binding of satellite groups is expected to be weaker due to their wider separations, allowing them to better phase-mix.

It is known that dynamical models are in general most sensitive to the gravity near the median radius of the tracer sample~\citep{oPDF}. One may wonder if the different performances of the different tracers can be attributed to them sampling different parts of the halo. It has been shown that the different spatial distributions of stars and dark matter only play a minor role in explaining their different systematics~\citep{oPDFII,Wang17}. Moreover, this effect is not relevant for the comparison between satellites and dark matter, because satellites trace the spatial distribution of dark matter fairly well~\citep{Han16}. We thus conclude that the different systematics between stars and satellites are mainly a result of their different dynamical states. 



\section{Conclusions and Implications}
We find that satellite galaxies are close to dark matter particles in their dynamical state, both of which are much better steady-state tracers than halo stars. This can be understood as satellite subhaloes are nearly unbiased phase space tracers of dark matter particles, while halo stars remain highly phase-correlated after getting stripped from their progenitors. Given that the effective number of phase-independent particles in halo stars is $\sim 40$, we expect the halo mass can be better constrained using $> 40$  satellite galaxies than using a much larger sample of halo stars. Taking the dark matter dynamical state as a reference, we expect a limiting accuracy of $\sim 20\%$ can be achieved from steady-state modelling of a few hundred satellites. A more accurate and practical method of modelling satellite dynamics will be presented in a forthcoming paper~\citep{Li19}.

\textit{This work was supported by JSPS Grant-in-Aid for Scientific Research JP17K14271.} 


\end{document}